\documentclass[twocolumn,preprintnumbers, superscriptaddress, reprint, longbibliography, pra]{revtex4-1}
\usepackage{graphicx}
\usepackage{amsmath}
\usepackage{amssymb}
\usepackage{MnSymbol}
\usepackage{times}
\usepackage{epstopdf}
\usepackage[colorlinks=true, linkcolor=blue, citecolor=blue, urlcolor=blue]{hyperref}
\usepackage{svg}

\begin{document}
	\title{Re-entrant Fulde-Ferrell-Larkin-Ovchinnikov State in Small-Sized Superconductors}
	
	\author{Tom Kim}
	\affiliation{Theoretical Division, T-4 and CNLS, Los Alamos National Laboratory, Los Alamos, New Mexico 87545, USA}
	\affiliation{School of Natural Sciences, University of California, Merced, CA 95343, USA}

	\author{Chih-Chun Chien}
	\affiliation{School of Natural Sciences, University of California, Merced, CA 95343, USA}
	
	\author{Shi-Zeng Lin}
\affiliation{Theoretical Division, T-4 and CNLS, Los Alamos National Laboratory, Los Alamos, New Mexico 87545, USA}

	\begin{abstract}
	We study the effect of a parallel magnetic field in a thin and small superconductor. The field suppresses superconductivity through Zeeman coupling while stabilizes the Fulde-Ferrell-Larkin-Ovchinnikov (FFLO) state at high fields before superconductivity is destroyed. When the spatial period of FFLO state is comparable to the size of the superconductor, there is a strong commensuration effect, which modifies the superconducting phase diagram. We investigate the FFLO state and the phase diagram in the presence of strong commensuration effect both for the $s$-wave and $d$-wave superconductors using the Bogoliubov de Gennes equation, Green function approach, and Ginzburg-Landau theory. We found that the superconducting phase diagram is strongly modulated. Interestingly, there is re-entrance of superconductivity upon increasing the magnetic field. The commensuration effect of the FFLO state can be used to detect the FFLO state in experiments.    
	\end{abstract}

	\maketitle
\section{Introduction}
Advance in nano-fabrication technique has opened many exciting possibilities to study superconductors with size comparable to the superconducting coherence length $\xi$. Generally, superconductivity is suppressed in small superconductors due to the level splitting of the electron states~\cite{anderson_theory_1959}. In those small superconductors, many new properties emerge due to the strong geometry confinement. For example, the surface energy barrier for a vortex to penetrate or exit the superconductor is highly asymmetric, resulting in a large hysteresis in magnetic field associated with vortex penetrating or exiting the superconductor~ \cite{PhysRevLett.12.14,PhysRevLett.101.167001,PhysRevLett.102.127005}. Other interesting phenomena originated from the geometry confinements include the paramagnetic Meissner effect \cite{geim_paramagnetic_1998}, symmetry-induced anti-vortex state \cite{chibotaru_symmetry-induced_2000}, Landau quantization of the superconducting macroscopic wave function under magnetic field \cite{geim_phase_1997}, and destruction of the global phase coherence when the flux in the superconductor is half integer of the flux quantum $\Phi_0=hc/2e$. \cite{liu_destruction_2001} In those studies, the geometry confinement on the orbital effect in superconductors under magnetic field has been investigated.        

Besides the coupling with the orbital motion of Cooper pairs, the magnetic field also couples to the spin of a Cooper pair through the Zeeman coupling. In type II superconductors, the magnetic field induces Abrikosov vortex lattice~\cite{Tinkham} through the orbital coupling. At high magnetic field, vortex cores start to overlap and the superconductivity is destroyed at a threshold field. This defines an orbital limited upper critical field $H_{\mathrm{orb}}=\Phi_0/2\pi\xi^2$. The Zeeman coupling of Cooper pairs also suppresses superconductivity with singlet pairing. By equating the superconducting condensation energy to the energy gain if the spins of a Cooper pair are fully polarized by magnetic field, one can define another upper critical field $H_p=\sqrt{2}\Delta/g\mu_B$, which is called the Chandrasekhar-Clogston  limit \cite{PhysRevLett.9.266,doi:10.1063/1.1777362}. Here $\Delta$ is the zero temperature superconducting energy gap, $\mu_B$ is the Bohr magneton and $g$ is the electron spin $g$ factor. In most superconductors, both the orbital effect and Zeeman effect are at work to suppress superconductivity by magnetic field. When $H_p\gg H_{\mathrm{orb}}$, the upper critical field is limited by orbital pairing breaking effect; while in the opposite limit when $H_{\mathrm{orb}}\gg H_{p}$, it is limited by Pauli pair breaking effect.

In superconductors with $H_{orb}\gg H_{p}$, also known as Pauli limited superconductors, Fulde and Ferrel \cite{PhysRev.135.A550} and, independently, Larkin and Ovchinnikov \cite{Sov.Phys.JETP20762} predicted a new superconducting phase with spatially modulated superconducting order parameter. In their honor, this state is now known as the FFLO state. The period of the modulation is comparable to $\xi$. In this way, the system can gain superconducting condensation energy in the region with nonzero superconducting order parameter and gain Zeeman energy when the superconducting order parameter vanishes. The phase boundary in the thermodynamic limit has been calculated \cite{PhysRevB.50.12760,PhysRevB.71.214504,shimahara_structure_1998}. Using the Bogoliubov-de Gennes (BdG) equation for a finite system, the order parameter, local density of states ~\cite{PhysRevLett.96.117006,0295-5075-86-4-47004}, and phase boundary also have been calculated~\cite{PhysRevB.78.054501,1367-2630-20-6-063001}.  Several families of superconductors have been identified as candidates for the realization of the FFLO state with encouraging experimental evidence \cite{PhysRevLett.91.187004,PhysRevLett.97.227002,PhysRevLett.99.187002,PhysRevLett.107.087002,PhysRevLett.109.027003,mayaffre_evidence_2014,PhysRevLett.116.067003,PhysRevLett.121.157004,PhysRevLett.119.217002,PhysRevB.85.174530,PhysRevB.97.144505}.
However, a conclusive experimental detection of the FFLO state in superconductors  remains a challenge ~\cite{,doi:10.1143/JPSJ.76.051005}. Recently, ultracold atomic Fermi gases in optical lattices provide a new platform to detect and characterize the FFLO state~\cite{Mizushima05,Koponen07,Parish07,Loh10,Liao10,Cai11}.

The thin and small superconductors provide a suitable platform to investigate the FFLO state. In thin superconductors, the orbital pair-breaking effect can be minimized by applying a magnetic field parallel to the superconductors. When the FFLO state is induced by the magnetic field, one would also expect a strong commensuration effect between the FFLO state and the geometry of the superconductor, hence resulting in modulated superconducting phase diagrams with superconductivity being suppressed (enhanced) when the size of the superconductor is incommensurate (commensurate) with the period of the FFLO state. The modulation of the phase diagram due to the commensuration effect can serve as experimental evidence of the elusive FFLO state. We remark that the experimental setups in Refs.~\onlinecite{PhysRevLett.101.167001,PhysRevLett.102.127005,PhysRevLett.117.116802} can be readily adapted to study the FFLO state in nano-sized superconductors~\cite{DirtySCNote}. 

In this work, we study the effect from geometric confinement on a small Pauli-limited superconductor under a magnetic field. We employ direct numerical calculations of the BdG equation for a tight-binding BCS Hamiltonian with $s$-wave or $d$-wave pairing interaction to demonstrate the generality of the commensuration effect. In addition, we compute the phase boundary between the superconducting and normal states using the Green function approach~\cite{Fetter_book}, which shows a good agreement with the BdG prediction. We also provide a more intuitive understanding based on the phenomenological Ginsburg-Landau approach~\cite{abrikosov1963methods,BUZDIN1997341}.  Based on these complementary methods, we find that the temperature-magnetic field phase diagram is strongly modulated due to the geometry confinement. Interestingly, there exist superconducting pockets separated by the normal region in the phase diagrams. As a consequence, an increasing magnetic field can induce a transition from the normal to superconducting phases due to the commensuration effect.

The reminder of the paper is organized as follows.  In Sec. \ref{sec2}, we detail
our model. The Bogoliubov de-Gennes equation is derived and solved numerically in Sec. \ref{sec3}. In Sec. \ref{sec4}, we calculate the phase boundary using the Green function approach. The temperature-magnetic field phase diagram, which is the main result of the present work, is presented in Sec. \ref{sec5}. The commensuration effect of the FFLO state based on the Ginzburg-Landau free energy functional is studied in Sec. \ref{sec6}. The paper is concluded by a summary in Sec. \ref{sec7}.

\section{Model}\label{sec2}
To demonstrate the  commensuration effect between the FFLO state and system size, we consider the mean-field tight binding Bardeen-Cooper-Schrieffer (BCS) Hamiltonian defined on a square lattice as an example. The generic features due to the geometric confinement are very general and are expect to be valid for more realistic model Hamiltonians. The Hamiltonian of our system reads
	\begin{eqnarray}
    \mathcal{H}=\sum _{ ij,\sigma   }  c _{i\sigma}^{\dagger} h_{ij,\sigma}  c _{j\sigma }
		+
		\sum_{ij}\left(\Delta _{ij} c ^{\dagger }_{i+ } c^{\dagger }_{j- }
		+
		\Delta _{ij}^* c_{j-}c _{i+ } \right),
	\end{eqnarray}
	where  $h_{ij,\sigma}=-t_{ij}-\mu_\sigma\delta_{ij}$. Here, $t_{ij}$ is the hopping coefficient from site $j$ to site $i$. We choose $t_{ij}=t$ for the nearest neighbor pairs and $t_{ij}=0$ otherwise. $\sigma=+$ for spin up and $\sigma=-$ for spin down.
    We consider the grand canonical ensemble and combine the chemical potentials $\mu_{\pm}=\mu\pm g\mu_B H/2$ with $\mu=(\mu_+ + \mu_-)/2$ and $h=(\mu_+ - \mu_-)/2=g \mu_B H/2$, where  $H$ denotes the magnetic field.  We consider a setup with a magnetic field parallel to the thin sample, therefore the orbital coupling of electrons to the gauge field is absent. The spin-orbit coupling is neglected, and the spin quantization axis is defined along the field direction in the following calculations. We consider the clean limit which is known to favor the FFLO state, and neglect the effect of impurities.
    
    The gap function, which also represents the superconducting order parameter, is
	\begin{eqnarray}
		\Delta_{ij}
		=\frac{1}{2}V_{ij}\left(\left\langle c_{i+
        } c _{j- }\right\rangle -\left\langle c _{i- }c_{j+ }\right\rangle \right),
	\end{eqnarray}
where $\langle A \rangle$ denotes the thermal expectation value of $A$ and $V_{ij}$ is the effective attraction between the electrons. The superconducting pairing symmetry depends on the functional form of $V_{ij}$. For instance, $V_{ij}=V\delta(i-j)$ stabilizes the $s$-wave pairing symmetry, so the order parameter is an onsite quantity. On the other hand, $V_{ij}=V \delta(i-j+\mathbf{r})$ favors the $d$-wave pairing symmetry, where $\mathbf{r}$ is a primitive vector. For the $d$-wave paring interaction, we use the following definition~\cite{PhysRevB.59.3353} 
\begin{eqnarray}
\Delta^{x}_i&=&\Delta_{i,i+\hat{x}},~~
\Delta^{y}_i=\Delta_{i,i+\hat{y}},\nonumber\\
\Delta^{s}_i
&=&\left(\Delta_{i,i+\hat{x}}+\Delta_{i,i-\hat{x}}+\Delta_{i,i+\hat{y}}+\Delta_{i,i-\hat{y}}\right)/4,\nonumber\\
\Delta^{d}_i
&=&\left(\Delta_{i,i+\hat{x}}+\Delta_{i,i-\hat{x}}-\Delta_{i,i+\hat{y}}-\Delta_{i,i-\hat{y}}\right)/4,
\label{eq:Delta_definition}
\end{eqnarray}
where $\hat{x}$ and $\hat{y}$ are the two primitive vectors. The $s$-wave component $\Delta^{s}_i$ is invariant while the $d$-wave component $\Delta^{d}_i$ changes sign under the $C_4$ rotation. We will consider open boundary condition in the following calculations.

\section{Bogoliubov de Gennes equation}\label{sec3}
	We use the Bogoliubov canonical transformation to diagonalize the grand-canonical Hamiltonian~\cite{JXZhuBook}. By introducing the quasi-particle operator $\gamma_n$, the electron operators can be written as a canonical transformation
	\begin{eqnarray}
		c_{i\sigma}=\sum_n\left(u_{i\sigma}^n\gamma_n-\sigma v_{i\sigma}^{n*}\gamma_n^{\dagger}\right),
		c_{i\sigma}^\dagger=\sum_n\left(u_{i\sigma}^{n*}\gamma_n^{\dagger}-\sigma v_{i\sigma}^{n}\gamma_n\right).
	\end{eqnarray}
The quasiparticle operators satisfy the anti-commutation relations
	\begin{eqnarray}
		\{\gamma_n,\gamma_m^\dagger\}=\delta_{nm},\  \{\gamma_n,\gamma_m\}=\{\gamma_n^\dagger,\gamma_m^\dagger\}=0,
	\end{eqnarray}
The diagonalized Hamiltonian has the following form
	\begin{eqnarray}
		\mathcal{H}=\sum_n E_n \gamma_n^\dagger \gamma_n +E_g,
	\end{eqnarray}
where $E_g$ is the ground state energy.
	Using the canonical transformation and the anti-commutation relations
	\begin{eqnarray}
		\{\mathcal{H},\gamma_n\}=E_n\gamma_n,\ \{\mathcal{H},\gamma_n^\dagger\}=-E_n\gamma_n^\dagger,
	\end{eqnarray}
we obtain the equation
	\begin{eqnarray}
		\sum_j M_{ij}\phi_j=E_n\phi_i,
	\end{eqnarray}
	where
	\begin{eqnarray}
		{\displaystyle M_{ij} 
			=\left({\begin{array}{rrrr}h_{ij,+}&0&0&\Delta_{ij}\\0&h_{ij,-}&\Delta_{ji} &0\\0 &\Delta_{ij}^* &-h_{ij,+} &0 \\\Delta_{ji}^*&0&0 &-h_{ij,-}\end{array}}\right)},~~
			\phi_i^n=\left(\begin{matrix}u_{i+}^n\\u_{i-}^n\\v_{i+}^n\\v_{i-}^n\end{matrix}\right).
	\end{eqnarray}
	Since the Bogoliubov-de Gennes (BdG) equation is in a block-diagonalized form when the spin-orbit coupling is absent, we can decompose it into two sets of coupled equations:
	\begin{eqnarray}
		E_{\tilde{n}1}\left(\begin{matrix}u_{i-}^{\tilde{n}1}\\v_{i+}^{\tilde{n}1}\end{matrix}\right)
		=
		\sum_j
		\left(\begin{matrix}h_{ij,+}^{\tilde{n}1}&\Delta_{ij}\\
			\Delta_{ij}^*&-h_{ij,-}\end{matrix}\right)
		\left(\begin{matrix}u_{j-}^{\tilde{n}1}\\v_{j+}^{\tilde{n}1}\end{matrix}\right),
		\label{eq:BdG_discrete}
	\end{eqnarray}
	and
	\begin{eqnarray}
		E_{\tilde{n}2}\left(\begin{matrix}v_{i-}^{\tilde{n}2}\\u_{i+}^{\tilde{n}2}\end{matrix}\right)
		=
		\sum_j
		\left(\begin{matrix}-h_{ij,+}^{\tilde{n}1}&\Delta_{ij}^*\\
			\Delta_{ij}&h_{ij,-}\end{matrix}\right)
		\left(\begin{matrix}v_{j-}^{\tilde{n}2}\\u_{j+}^{\tilde{n}2}\end{matrix}\right).
	\end{eqnarray}
	The first equation can be rewritten as
	\begin{eqnarray}
		-E_{\tilde{n}1}\left(\begin{matrix}v_{i-}^{\tilde{n}1*}\\-u_{i+}^{\tilde{n}1*}\end{matrix}\right)
		=
		\sum_j
		\left(\begin{matrix}-h_{ij,-}^{\tilde{n}1}&\Delta_{ij}\\
			\Delta_{ij}^*&h_{ij,+}\end{matrix}\right)
		\left(\begin{matrix}v_{j-}^{\tilde{n}1*}\\-u_{j+}^{\tilde{n}1*}\end{matrix}\right).
	\end{eqnarray}
	Comparing to the second equation, we obtain the equivalence
	\begin{eqnarray}
		\left(\begin{matrix}v_{i-}^{\tilde{n}2}\\u_{i+}^{\tilde{n}2}\end{matrix}\right)
		=
		\left(\begin{matrix}v_{i-}^{\tilde{n}1*}\\-u_{i+}^{\tilde{n}1*}\end{matrix}\right),
	\end{eqnarray}
	and $E_{\tilde{n}2}=-E_{\tilde{n}1}$. This symmetry allows us to rewrite the canonical transformation using only the positive-energy states:
	\begin{eqnarray}
		&&c_{i+}=\sum_{\tilde{n}}^{\prime}\left(u_{i+}^{\tilde{n}1}\gamma_{\tilde{n}1}- v_{i+}^{\tilde{n}2*}\gamma_{\tilde{n}2}^{\dagger}\right),
		c_{i+}^\dagger=\sum_{\tilde{n}}^{\prime}\left(u_{i+}^{\tilde{n}1*}\gamma_{\tilde{n}1}^\dagger- v_{i+}^{\tilde{n}2}\gamma_{\tilde{n}2}\right),\nonumber\\
		&&c_{i-}=\sum_{\tilde{n}}^{\prime}\left(u_{i-}^{\tilde{n}2}\gamma_{\tilde{n}2}+ v_{i-}^{\tilde{n}1*}\gamma_{\tilde{n}1}^{\dagger}\right),
		c_{i-}=\sum_{\tilde{n}}^{\prime}\left(u_{i-}^{\tilde{n}2*}\gamma_{\tilde{n}2}^\dagger+ v_{i-}^{\tilde{n}1}\gamma_{\tilde{n}1}\right). \nonumber \\
	\end{eqnarray}
Here $\sum^\prime$ denotes the summation over only the positive energy states.
The diagonalized Hamiltonian is
\begin{eqnarray}\label{eq16}
\mathcal{H}=\sum _{\tilde{n}} ^\prime E_{\tilde{n}} \gamma _{\tilde{n}}^{\dagger} \gamma _{\tilde{n}}+\sum _{i,j} \frac{\left| \Delta_{ij}\right| ^2}{V_{ij}},
\end{eqnarray}
where $\tilde{n}=n1,n2$ comes from the Bogoliubov transformation and we drop the constant ground-state energy.

	Using the canonical transformation, the gap function becomes 
	\begin{eqnarray}
		\Delta_{ij}=\frac{V_{ij}}{4}\sum_{\tilde{n}}\left(
		u_{i+}^{\tilde{n}1} v_{j-}^{\tilde{n}1*}
		+
		u_{j+}^{\tilde{n}1} v_{i-}^{\tilde{n}1*}
		\right)
		\tanh\left(\frac{E_{\tilde{n}}}{2 k_B T}\right).
		\label{eq:order_parameter}
	\end{eqnarray}
Here  $k_B$ is the Boltzmann constant.
Numerically, the order parameter can be found by the following procedure: Let $\Delta_{ij,\nu}$ be the order parameter of the $\nu$-th iteration. With an initial guess of the order parameter, $\Delta_{ij,0}$, we calculate the eigenvectors and eigenvalues using Eq.~\eqref{eq:BdG_discrete}. With the eigenvalues and eigenvectors, we then calculate the order parameter for the next iteration, $\Delta_{ij,1}$, using Eq. \eqref{eq:order_parameter}. We repeat the process until the order parameter converges, i.e. until $\sum_{ij} | |\Delta_{ij,\nu}|-|\Delta_{ij,\nu+1}| |/t$ is smaller than a preset tolerance value of $10^{-16}$.

Depending on the initial values of the BdG equation, the system may be trapped in a metastable state rather than converge to the thermodynamically stable state. In order to find the thermodynamically stable state and the phase boundary between one superconducting state and another,  we need to compare the free energy, which can be obtained as follows.
The entropy of the Bogoliubov quasiparticle in Eq. \eqref{eq16} is~\cite{ketterson1999superconductivity}
\begin{eqnarray}
S =
\sum_i \beta  E_i f\left(E_i\right)\ln \left(e^{-\beta E_i}+1\right),
\end{eqnarray}
where $f(x)=(\exp(\beta x)+1)^{-1}$ is the Fermi function with $\beta=(k_B T)^{-1}$.
From the relation $\mathcal{F}=\langle \mathcal{H}\rangle -S T$ we obtain the free energy.
\begin{eqnarray}
\mathcal{F}=-\frac{1}{\beta}\sum _{\tilde{n}=n1,n2}^\prime \ln \left(e^{-\beta  E_{\tilde{n}}}+1\right)+\sum _{i,j} \frac{\left| \Delta_{ij}\right| ^2}{V_{ij}}.
\end{eqnarray}


	\section{Green function approach}\label{sec4}
	The order parameter $\Delta$  is in general not uniform, and Bogoliubov de Gennes equation can be used to obtain its profile. However, a system of size $N$ requires the diagonalization of $2N$ by $2N$ matrices, so the size of the system will be limited in the numerical calculations. If we are interested only in finding the phase boundary between the superconducting and normal phases, where the order parameter approaches zero, we can obtain a semi-analytic solution by expanding the Green function around the phase boundary where the gap vanishes.
	
	 The Green function method for systems in equilibrium is usually cast in the imaginary time formalism \cite{Fetter_book}. Following Ref.~\cite{Fetter_book}, we construct the thermal operator $c_i(\tau)=e^{k_B \tau}c_i e^{-k_B \tau}$ with the imaginary time $\tau$. The imaginary time evolution equations are 
	\begin{eqnarray}
	\hbar \partial _{\tau }c _{i\pm }(\tau)
	=\sum_{j }\left(  -h_{ij,\pm }c_{j\pm }( \tau) \mp \Delta_{ij}c^{\dagger }_{j\mp }(\tau )
	\right),
	\end{eqnarray}
	\begin{eqnarray}
	\hbar \partial _{\tau }c _{i\pm}^\dagger(\tau)
	=\sum_{j }  \left(
	h_{ij,\pm }  c_{j\pm }^\dagger( \tau)\pm \Delta_{ij}c_{j\mp }(\tau )
	\right).
	\end{eqnarray}
	We define the imaginary time ordered Green functions
	\begin{eqnarray}
		G_{\alpha, \beta }\left(x_1\tau _1,x_2\tau _2\right)=-\left\langle \mathcal{T_\tau} \left\{c _{x_1\alpha }\left(\tau _1\right) c ^{\dagger }_{x_2 \beta }\left(\tau _2\right)\right\}\right\rangle,
	\end{eqnarray}
	\begin{eqnarray}
		F_{\alpha, \beta }\left(x_1\tau _1 ,x_2\tau _2 \right)=-\left\langle \mathcal{T_\tau} \left\{c _{x_1\alpha }\left(\tau _1 \right) c _{x_2\beta }\left(\tau _2 \right)\right\}\right\rangle,
	\end{eqnarray}
	\begin{eqnarray}
		F^{\dagger }_{\alpha ,\beta}\left(x_1\tau _1 ,x_2\tau _2 \right)=\left\langle \mathcal{T_\tau} \left\{c ^{\dagger }_{x_1\alpha }\left(\tau _1 \right) c ^{\dagger }_{x_2\beta }\left(\tau _2 \right)\right\}\right\rangle.
	\end{eqnarray}
	Here $T_\tau$ denotes imaginary time ordering and $\alpha,\beta=+,-$ .
    
By using the equations for the fermion operators, we obtain the equations for the Green function. 
	\begin{eqnarray}
		\left(-\hbar \partial _{\tau _1}-\sum_{x}h_{x_1 x,\pm}\right)
		G_{\pm\pm }(x\tau _1,x_2\tau _2 )\nonumber\\
		\pm
		\sum_x \Delta_{x_1,x} F^{\dagger }_{\mp \pm }(x\tau _1 ,x_2\tau _2)\nonumber\\
		=
		\hbar  \delta \left(\tau _1-\tau _2\right) \delta(x_1- x_2),
		\label{eq:greens_function_discrete_equation_of_motion1}
	\end{eqnarray}
	\begin{eqnarray}
		\left(-\hbar \partial _{\tau _1}+\sum_{x}h_{x_1 x,\pm}\right)F^{\dagger }_{\mp\pm }(x\tau _1 ,x_2\tau _2 )\nonumber\\
		\pm
		\sum_{x} \Delta^*_{x_1,x}G_{\pm\pm} (x\tau _1 ,x_2\tau _2 ) =0.
		\label{eq:greens_function_discrete_equation_of_motion2}
	\end{eqnarray}
Using the Green function method to solve for a general case can be difficult. However, the method can be used to find the equation determining the normal-superconducting phase boundary where $\Delta\rightarrow 0$.
	
When $\Delta=0$, we obtain the normal state Green functions using the properties of tridiagonal Toeplitz matrices summarized in Appendix~\ref{App:A}. Explicitly, 
	\begin{eqnarray}
		G_{0\pm\pm}(x_1\tau _1 ,x_2\tau _2 )
&=&\sum _{k,\omega_n } u_{k \omega_n}\left(x_1 \tau_1\right) u_{k \omega_n}^*\left(x_2 \tau_2\right)\nonumber\\
		&&\times
		\frac{\hbar }{i \omega_n  \hbar -\xi _{\pm}(k)}, \\
u_{k \omega_n}\left(x \tau\right)&=&\frac{2}{\sqrt{(N_x+1)(N_y+1)\hbar\beta}}\sin\left(\frac{\pi x^{(1)} k_x}{N_x+1}\right)\times \nonumber \\
& &\sin\left(\frac{\pi x^{(2)} k_y}{N_y+1}\right)\exp\left(-i\tau\omega_n\right). \nonumber
        \label{eq:normal_greens_function}
	\end{eqnarray}
	where $\xi _{\pm }(k)=-2\left(\cos(\frac{\pi k_x}{N_x+1})+\cos(\frac{\pi k_y}{N_y+1})\right)-\mu\mp h$ for the two-dimensional rectangular lattice. Here $\omega_n=\left(2n+1\right)\pi/\beta\hbar$ is the fermionic Matsubara frequency and $x=(x^{(1)},x^{(2)})$.
	The solutions for $G$ and $F^\dagger$ are
	\begin{eqnarray}
		&&G_{\pm\pm}(x_1\tau _1 ,x_2\tau _2 )\nonumber\\
		&&=
		G_{0\pm\pm}(x_1\tau _1 ,x_2\tau _2 )
		\mp\frac{1}{\hbar}\sum_{x_3,x_4}\int d\tau \nonumber\\
		&&\times G_{0\pm\pm }(x_1\tau _1,x_3\tau )\Delta ^*_{x_3,x_4}F_{\mp\pm}^\dagger(x_4\tau  ,x_2\tau _2 ),
	\end{eqnarray}
	\begin{eqnarray}
		&&F^{\dagger }_{\mp\pm}(x_1\tau _1 ,x_2\tau _2 )\nonumber\\
		&&=\pm\frac{1}{\hbar}\sum_{x_3,x_4}\int d\tau \nonumber\\
		&&\times G^*_{0\mp\mp }(x_3\tau,x_1\tau _1 )\Delta ^*_{x_3,x_4}G_{\pm\pm }(x_4\tau  ,x_2\tau _2 ),
	\end{eqnarray}
	which can be verified by plugging it back to Eq.~\eqref{eq:greens_function_discrete_equation_of_motion1} and ~\eqref{eq:greens_function_discrete_equation_of_motion2}. Using the definition of the order parameter, $\Delta ^*_{x_1,x_2}=V\left(x_1-x_2\right)\left(F^{\dagger }_{-+}(\tau _1 x_1,\tau _1 x_2)-F^{\dagger }_{+-}(\tau _1 x_1,\tau _1 x_2)\right)/2$, we obtain the expansion to the lowest order in $\Delta$,
	\begin{eqnarray}
		&&
        \Delta ^*_{x_1,x_2}
		=
		\frac{1}{2\hbar}\sum_{x_3,x_4} \int d\tau 
		V\left(x_1-x_2\right)\nonumber\\
		&&
        \times 
        \left[
        G^*_{0-- }(\tau  x_3,\tau _1 x_1)\Delta ^*_{x_3,x_4}G_{0++ }(\tau  x_4,\tau _1 x_2)\right.\nonumber\\
        &&
        +
        \left.G^*_{0++}(\tau  x_3,\tau _1 x_1)\Delta ^*_{x_3,x_4}G_{0--}(\tau  x_4,\tau _1 x_2)\right].
\label{eq:self_consistent_equation}
	\end{eqnarray}
The second term inside the square bracket is identical to the first term except the up spin and down spins are switched. This equation determines the phase boundary between the superconducting and normal states as $\Delta$ vanishes.

\subsection{$s$-wave pairing interaction}
	For the system with $s$-wave pairing interactions, we assume
	\begin{eqnarray}
		V\left(x_1-x_2\right)=V \delta \left(x_1-x_2\right).
	\end{eqnarray}
	The order parameter can be written as
	\begin{eqnarray}
		\Delta ^*_{x_1,x_2}
		= \delta (x_1-x_2) \Delta^*_{x_1}.
	\end{eqnarray}
	Summing over $x_2$, the self consistent equation~\eqref{eq:self_consistent_equation} at the phase boundary becomes
	\begin{eqnarray}
        \Delta ^*_{x_1}
        &&
		=
		\frac{V}{2\hbar}\sum_{x_3}\int d\tau \left[ G^*{}_{0-- }(\tau  x_3,\tau _1 x_1)\Delta ^*_{x_3}G_{0++ }(\tau  x_3,\tau _1 x_1)\right.\nonumber\\
        &&
        +
        \left.\left(+\sigma\leftrightarrow-\sigma\right)\right],
		\label{eq:self_consistent_equation_swave}
	\end{eqnarray}
which is in the form of $\sum_{x_3}\Delta^*_{x_3}\left[ A(x_3,x_1)-\delta(x_3,x_1) \right]=0$, where $A(  x_3,x_1)=(V/2\hbar)\int d\tau \left[G^*_{0-- }(\tau  x_3,\tau _1 x_1)G_{0++}(\tau  x_3,\tau _1 x_1)+\left(\sigma\leftrightarrow-\sigma\right)\right]$. 

\begin{figure}[b]
        \includegraphics[width=1\linewidth]{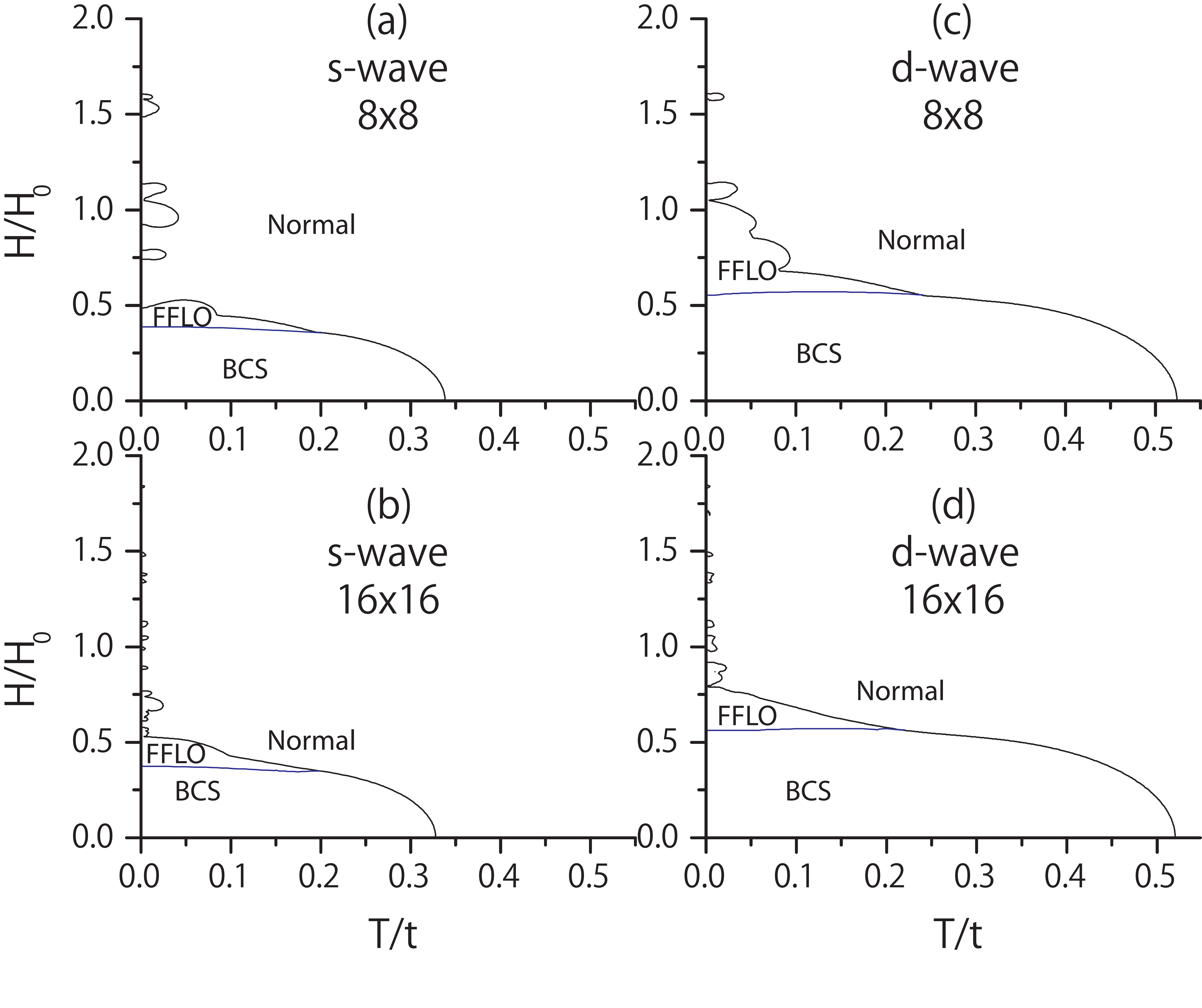}
        \caption{Phase diagrams of thin and small superconductors at $\mu=-0.4$ and $V/t=2.5$. The pockets near the vertical axis are the reentrant FFLO state. The superconducting-normal phase boundary can be obtained from the BdG equation or the Green function method, and their results agree with each other. The BCS-FFLO phase boundary, on the other hand, is obtained from the BdG equation. The results are obtained both for the $s$ wave (left column) and $d$ wave (right column) superconductors with system size $(L_x, L_y) = (8,8)$ (top row) and $(16,16)$ (bottom row). Here $H_0=2t/g \mu_B$.}
        \label{fig:phase_boundary_combined}
\end{figure}

\subsection{$d$-wave pairing interaction}
For the system with $d$-wave pairing interaction, we assume
\begin{eqnarray}
V\left(x_1-x_2\right)=V \sum _r \delta \left(r+x_1-x_2\right),
\end{eqnarray}
where $r=\left\{r_x \hat{x},-r_x \hat{x},r_y \hat{y},-r_y \hat{y}\right\}$ is the collection of the primitive vectors to the nearest neighbors.
The order parameter can be written as 
\begin{eqnarray}
\Delta ^*_{x_1,x_2}
&=&
\sum _r \delta \left(r+x_1-x_2\right) \Delta ^{r*}_{x_1},\nonumber\\
\Delta ^{r*}_{x_1}&=&\Delta ^*_{x_1,x_1+r}.
\end{eqnarray}
Here $\Delta ^{-r}_{x}=\Delta^r_{x-r}$ since $\Delta _{x_1,x_2}=\Delta_{x_2,x_1}$. The equation for the phase boundary  becomes
\begin{eqnarray}
&&\sum _{r_1} \delta \left(r_1+x_1-x_2\right) \Delta ^{r_1*}_{x_1}\nonumber\\
&&=\frac{1}{\hbar}\sum _{r_1,r_2,x_3,x4} \int d\tau V\delta \left(r_1+x_1-x_2\right)\nonumber\\
&&
\times\left[G^*_{0-- }\left(\tau  x_3,\tau _1 x_1\right)\Delta ^{r_2*}_{x_3}\right.\nonumber\\
&&\times
\left.
\delta \left(r_2+x_3-x_4\right)G_{0++ }\left(\tau  x_4,\tau _1 x_2\right)\right)\nonumber\\
&&
\left.+G^*{}_{0++}\left(\tau  x_3,\tau _1 x_1\right)\Delta ^{r_2*}_{x_3}\right.\nonumber\\
&&\times
\left.\delta \left(r_2+x_3-x_4\right)G_{0--}\left(\tau  x_4,\tau _1 x_2\right)\right].
\end{eqnarray}
The $\delta$ function is satisfied when $x_2=x_1+r_1$, and the equation becomes
\begin{eqnarray}
\Delta ^{r_1*}_{x_1}
&=&
\frac{1}{2\hbar}\sum _{r_2,x_3} \int d\tau V \left[G^*_{0-- }\left(\tau  x_3,\tau _1 x_1\right)\Delta ^{r_2*}_{x_3}\right.\nonumber\\
&&\times
\left. G_{0++}\left(\tau  \left(r_2+x_3\right),\tau _1 \left(r_1+x_1\right)\right)\right.\nonumber\\
&&\left.+\left(+\sigma \leftrightarrow -\sigma\right) \right],
\label{eq:self_consistent_equation_dwave}
\end{eqnarray}
which is in the form of $\sum_{r_2,x_3}\Delta^{r_2*}_{x_3}\left[A\left(r_2+x_3,r_1+x_1\right)-\delta(x_3,x_1)\delta(r_2,r_1)\right]=0$,  where $A\left(r_2+x_3,r_1+x_1\right)=\frac{V}{2\hbar}\int d\tau [G^*_{0-- }\left(\tau  x_3,\tau _1 x_1\right)G_{0++}\left(\tau  \left(r_2+x_3\right),\tau _1 \left(r_1+x_1\right)\right)+\left(+\sigma \leftrightarrow -\sigma\right)]$.
Techniques for calculating $A\left(r_2+x_3,r_1+x_1\right)$ can be found in Appendix~\ref{App:B}.

\section{Phase diagram}\label{sec5}
Our system can host the  normal, BCS, and FFLO states. The BCS state has uniform superconducting order parameter except in the region near the sample edges. In the following discussion, we refer to the superconducting state without any sign change of the order parameter as the BCS state, and those exhibiting sign changes as the FFLO state. 
 
To locate the phase boundary between the normal and superconducting states, we can use either Green function method or solve the BdG equation numerically. To use the Green function method, we solve the self-consistent equations~\eqref{eq:self_consistent_equation_swave} and ~\eqref{eq:self_consistent_equation_dwave}, which are of the form $(A-1)\Delta=0$ in matrix notation. It can be shown that $A$ approaches zero at high temperatures and strong magnetic field, so the only solution is $\Delta=0$. The eigenvalues of $A-1$ depend on temperature and magnetic field, and we need to find the eigenvalue that maximizes the superconducting region at a given temperature or field.  If we approach the normal-superconducting phase boundary from the normal state, one of the eigenvalues of $A-1$ approaches zero from the negative side. By locating the zeros of the largest eigenvalue of $A-1$, we find the phase boundary between the normal and superconducting states from the Green function method. We also use the BdG equation to locate the phase boundary by calculating where the order parameter becomes zero. The results from both methods agree well.

To find the BCS-FFLO phase boundary, the Green function method cannot be used since the derivation is not valid when the order parameter is not small, which is the case inside the superconducting regime. Therefore, we use the BdG equation to locate the BCS-FFLO phase boundary. By using different initial values of the order parameter, we obtain different metastable solutions from the BdG equation. At given values of temperature and magnetic field, we compare the free energies of the BCS state and the FFLO state and find the one with the lowest free energy. By locating where the free energies of the BCS and FFLO states are equal, we find the phase boundary of the first-order transition between the BCS and FFLO states.

Figure~\ref{fig:phase_boundary_combined} shows the phase boundary between the BCS, FFLO, and normal states for the $s$ wave and $d$ wave superconductors. The phase boundary between the FFLO and normal states is found by expanding the Green function, and the boundary between the BCS and FFLO states is found by solving the BdG equation and comparing the free energies of the FFLO and BCS states. The transition between the superconducting and normal states is of the second order, while the transition between different superconducting states is of the first order~\cite{Landau_SM1,doi:10.1143/JPSJ.76.051005}.

One anomalous behavior in the phase diagram is the re-entrance of superconductivity upon increasing field at low temperatures. On the phase diagrams shown in Fig.~\ref{fig:phase_boundary_combined}, one can see the small superconducting pockets separated by the normal phase at low temperatures, showing re-entrance of the FFLO state. The reason for the re-entrance of FFLO state is that the FFLO state has modulations of its order parameter. Thus, the system is stable if the modulation fits the finite system size. The characteristic length of the modulation is controlled by the magnetic field. If the modulation cannot fit the system size, the FFLO state can be suppressed. However, if at higher magnetic field the modulation can match the system size, the system can re-enter the FFLO state.  Note that the re-entrance mechanism here is different from the one discussed in Ref. \cite{PhysRevB.84.054518}, where the thermodynamic limit has been taken in a bilayer system.

\begin{figure}[t]
        \includegraphics[width=1\linewidth]{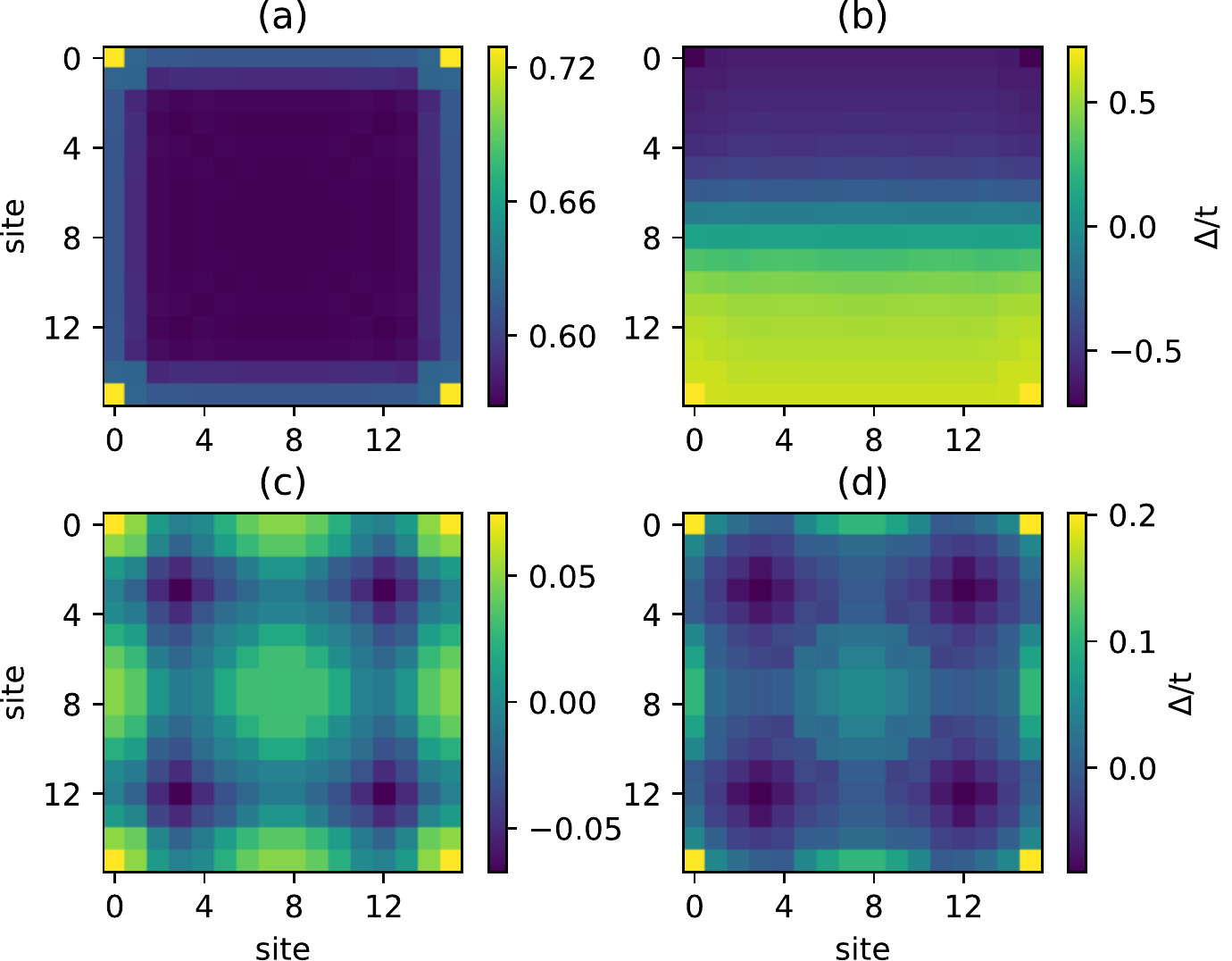}
        \caption{Profiles of the order parameter for the $s$-wave pairing interaction on a $16\times16$ lattice at $T=0$. Here $\mu/t=-0.4$ and $V/t=2.5$. (a) $H/H_0=0$, (b) $H/H_0=0.4$, (c) $H/H_0=0.55$, (d) $H/H_0=0.7$.}
        \label{fig:swave_16x16_gallery}
\end{figure}

\begin{figure}[t]
        \includegraphics[width=1\linewidth]{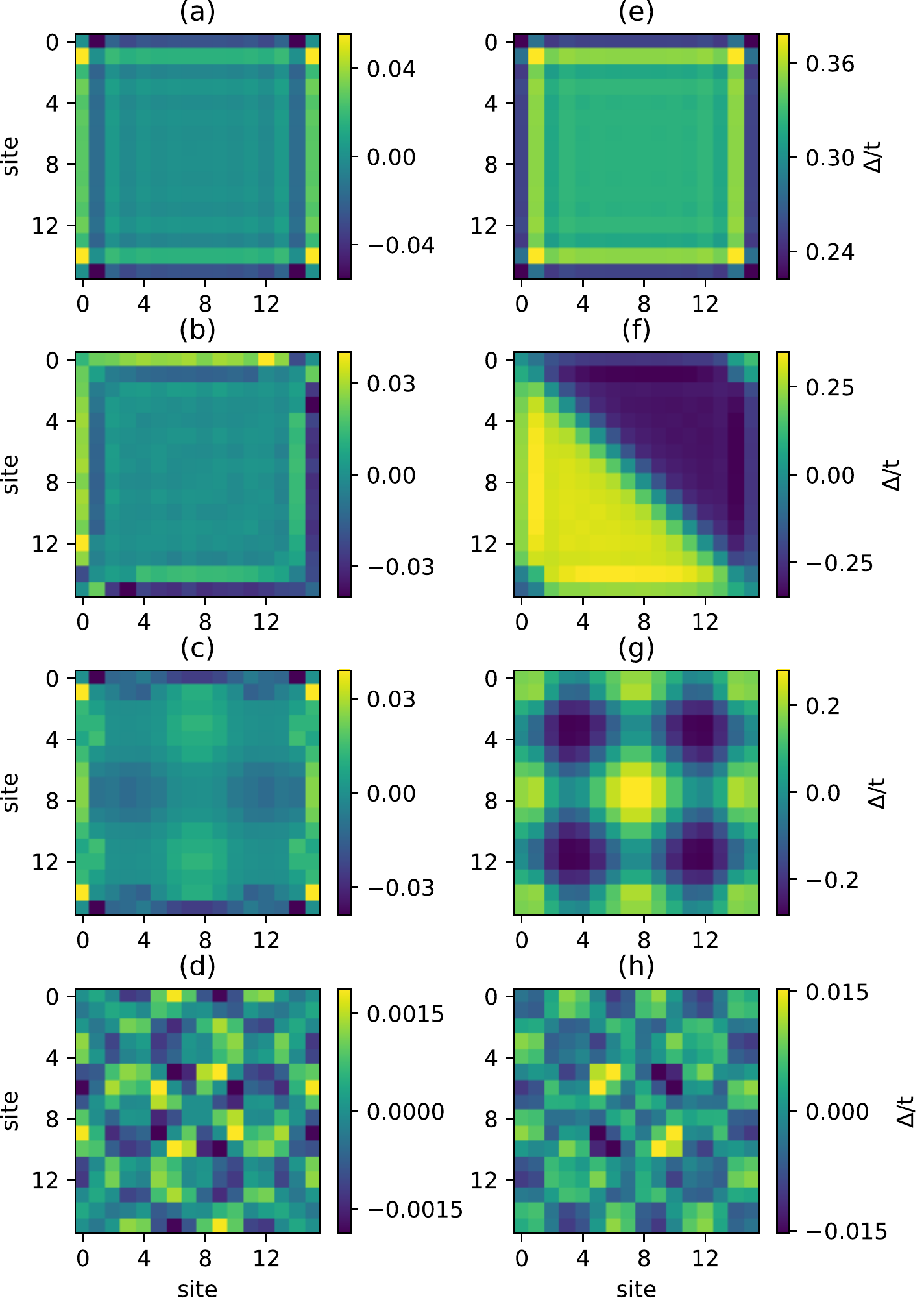}
        \caption{Profiles of the order parameter, defined in Eq.~\eqref{eq:Delta_definition}, of $d$-wave pairing interaction in a 16x16 system at $T=0$, $\mu/t=-0.4$, and $V/t=2.5$. The left and right columns show $\Delta^s$ and  $\Delta^d$, respectively.
        (a) and (e): $H/H_0=0$ in the BCS state, (b) and (f): $H/H_0=0.52$ in the FFLO state, (c) and (g): $H/H_0=0.65$ in the FFO state near the normal-superconducting phase boundary, (d) and (h): $H/H_0=1$ in the re-entrant FFLO sate due to the commensuration effect.}
        \label{fig:dwave_16x16_sd_gallery}
\end{figure}

Figure ~\ref{fig:swave_16x16_gallery} shows the typical spatial profiles of the order parameter from the BdG equation with $s$-wave pairing interaction at zero temperature. We used $\mu/t=-0.4$ and $V/t=2.5$ following Ref.~\cite{PhysRevLett.96.117006}. Because of the absence of translational invariance due to open boundary condition, the order parameter $\Delta$ is not uniform even when $H=0$, as shown in Fig. \ref{fig:swave_16x16_gallery} (a). At higher field when superconductivity still survives, $\Delta$ changes signs in space, which is a defining feature of the FFLO state. In Fig. \ref{fig:swave_16x16_gallery} (b), $\Delta$ is modulated along one of the spatial direction. As the magnetic field increases,  $\Delta$ starts to develop modulations in both $x$ and $y$ directions, as shown in Fig. \ref{fig:swave_16x16_gallery} (c) and (d).

The profiles of $\Delta^s$ and $\Delta^d$, defined in Eq.~\eqref{eq:Delta_definition}, are displayed in Fig.~\ref{fig:dwave_16x16_sd_gallery} for the $d$-wave pairing interaction. In the BCS state, a subdominant $s$-wave component is induced at the boundary, where the $C_4$ rotation symmetry is absent. The presence of the FFLO state also breaks the $C_4$ rotation symmetry, and a finite $s$-wave component is generated. Similar to those in Fig. ~\ref{fig:swave_16x16_gallery}, the FFLO states can have spatial modulations in the $x$ and/or $y$ directions at high fields.

\begin{figure}[t]
        \includegraphics[width=1\linewidth]{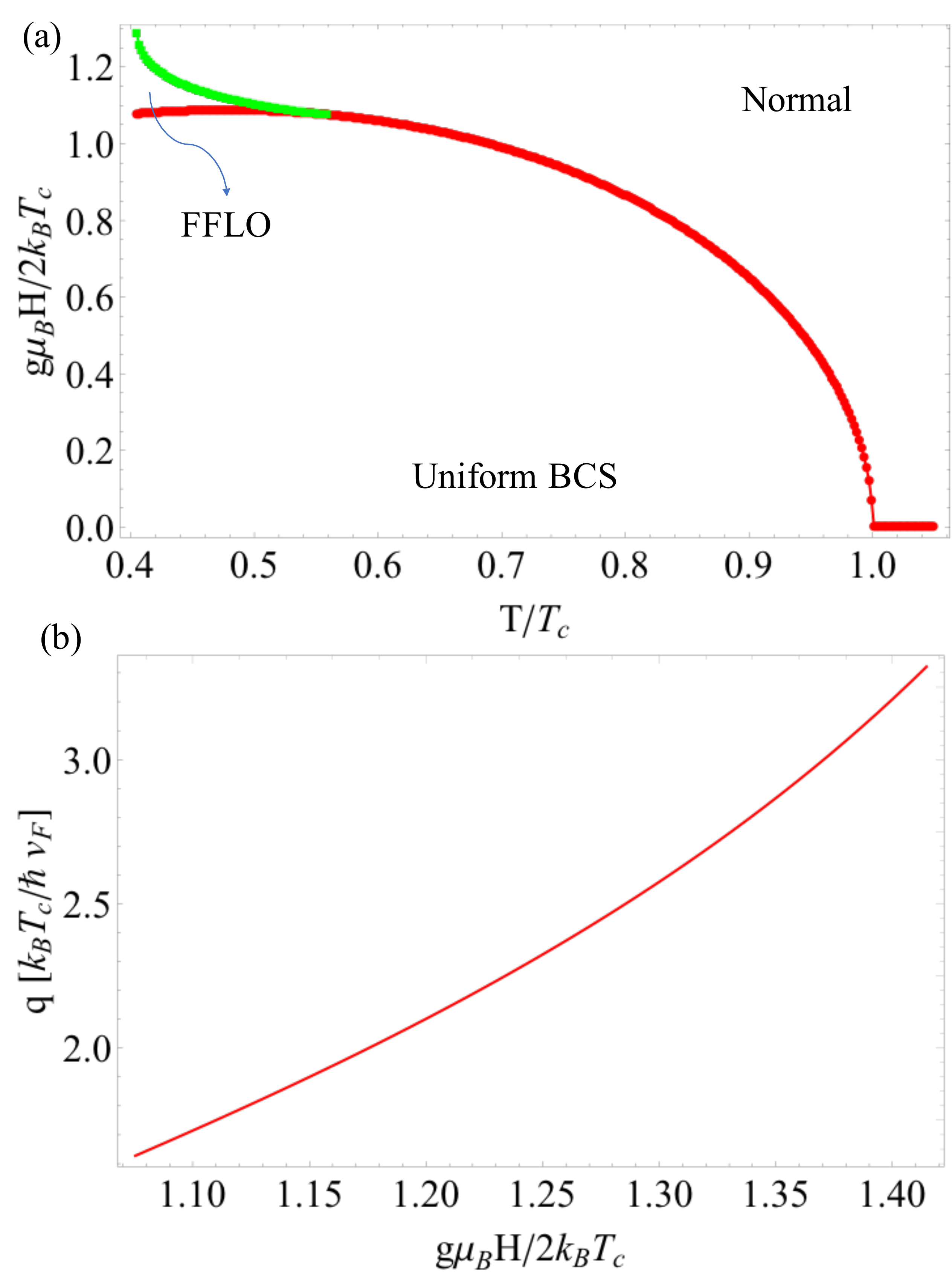}
        \caption{(a) Superconducting phase diagram and (b) FFLO wave vector versus $H$ obtained using the Ginzburg-Landau theory in 1D. The red line in (a) corresponds to the normal-superconducting phase boundary by assuming a BCS state with $q_0=0$. Here $T=0.4 T_c$ in (b).}
        \label{fGL1}
\end{figure}

\section{Ginzburg-Landau approach}\label{sec6}
Here we provide a more transparent understanding of the commensuration effect of the FFLO state based on the Ginzburg-Landau theory. The Ginzburg-Landau free energy function has been derived near the tricritical point in the $T$-$H$ phase diagram, where both the superconducting order parameter and wave vector of the FFLO state are small~\cite{BUZDIN1997341}.  The Ginzburg-Landau free energy functional, up to the order of $|\psi|^6$, can be written as 
\begin{align}
\mathcal{F}=\alpha |\psi|^2+\beta' |\nabla\psi |^2+\gamma|\psi|^4+\delta| \nabla ^2\psi| ^2+\mu |\psi ^2|| \nabla \psi|^2\\ \nonumber
+\eta \left[\left(\psi ^*\nabla \psi \right)^2+\left(\psi \nabla \psi ^*\right)^2\right]+\upsilon | \psi|^6.
\end{align}
Here $\psi$ is the order parameter, and $T$ and $H$ enter $\mathcal{F}$ through the coefficients. For a clear demonstration, we consider the 1D case. The analysis can be generalized to higher dimensions and the results are qualitatively similar. For instance, for a disk in 2D, the FFLO state close to the normal state phase boundary can be described by using the Bessel functions in the polar coordinates. 

First, let us calculate $T_c(H)$ as a function of the field $H$ in the thermodynamic limit $L\rightarrow \infty$, where $L$ is the system size. Close to $T_c$, we only keep the quadratic terms 
\[
\mathcal{F} \left(T\to T_c\right)=\alpha |\psi |^2+\beta' |\partial_x \psi |^2+\delta | \partial_x ^2\psi | ^2.
\]
For a parabolic dispersion of the electron band, the coefficients are \cite{BUZDIN1997341}
\[
\alpha =-\pi\left(K_1-K_1^0\right) N_0,\ \beta' =\frac{1}{4} \pi  K_3 N_0\hbar^2 v_F^2,\ \delta =\frac{-\pi}{16} K_5 N_0 \hbar^4 v_F^4
\]
\[
K_1=-\frac{\mathrm{Re}\left[\Psi(z)\right]}{\pi },\ K_3=\frac{-2 k_B T \mathrm{Re}\left[\Psi ^{(2)}(z)\right]}{2 (2 \pi  k_B T)^3},
\]
\[
K_5=\frac{-2 k_B T \mathrm{Re}\left(\Psi ^{(4)}(z\right))}{4! (2 \pi  k_B T)^5},\ z=\frac{1}{2}-\frac{i g \mu_B H}{4 \pi  k_B T},
\]
where $\Psi^{(n)}$ is the $n$-th derivative of the poly digamma function, $v_F$ is the Fermi velocity, and $N_0$ is the density of state at the Fermi surface. Here $K_1^0=-{\mathrm{Re}\left[\Psi  \left(\frac{1}{2}-\frac{i g\mu_B H_s}{4 \pi  k_B T}\right)\right]}/{\pi }$, where $H_s$ is the field corresponding to the second-order transition into the uniform superconducting state. Moreover, we have the following relation:
\begin{align}
\ln \frac{ T_c}{T}=\mathrm{Re}\left[-\frac{\Psi }{2}+\Psi  \left(\frac{1}{2}-\frac{i g\mu_B H_s}{4 \pi k_B T}\right)\right].
\end{align}
The FFLO solution is stable when $\beta'<0$ and $\delta>0$, and its order parameter can be written as $\psi =A \exp  (i q r)$ with a continuous $q$. The optimal $q$ is $q_{\mathrm{opt}}^2=-\frac{\beta' }{2 \delta }=\frac{2 K_3}{K_5 \hbar ^2v_F^2}$, and $T_c(H)$ is determined by $K_1-K_1^0=\frac{1}{4} K_3^2/ K_5$. The resulting phase diagram is shown in Fig. \ref{fGL1} (a). 

The Ginzburg-Landau energy functional $\mathcal{F}$ can only be derived rigorously near the normal-superconductor phase boundary. However, one can still obtain qualitative feature about the dependence of $q_{\mathrm{opt}}$ on $H$ deep inside the superconducting phase. The field dependence of $q_{\mathrm{opt}}$ at $T/T_c=0.4$ is displayed in Fig. \ref{fGL1} (b). The period of modulation is of the order of $\xi$. Moreover,  $q_{\mathrm{opt}}$ increases with $H$ because the separation between the Fermi surfaces of the up-spin and down-spin channels increases. The Cooper pair with opposite electron spins thus also has large momentum, which is just $q_{\mathrm{opt}}$ of the FFLO state.

Now we consider a superconducting wire of length $L$ subjected to the boundary condition $\partial_x\psi=0$ at both ends. The FFLO state is now described by $\psi =A \exp  (i q_n r)$, where $q_n =\frac{\pi  n}{L}$ with an integer $n$. The phase boundary is given by
\[
\alpha +\beta'  q_n^2+\delta  q_n^4=0.
\]
The optimal $q_n$ is the $q_n$ that maximizes the superconducting phase region. Close to the phase boundary, we can expand
\begin{align}\label{EqExp1}
\alpha -\frac{{\beta'}^2}{4 \delta }=-c_1 \Delta _H^2,\ q_{\mathrm{opt}}=c_2 \Delta _H+\bar{q},
\end{align}
with $\bar{q}$ the $q$ value at $H=H_c$ , $\Delta _H=\frac{H_c-H}{H_c}$, and $H_c$ the upper critical field for $L\rightarrow\infty$. The first (second) expression in Eq. \eqref{EqExp1} is even (odd) in $\Delta_H$ because the free energy (wave vector $q$) is even (odd) under time reversal operation, $H\rightarrow -H$. Then, the phase boundary is given by
\[
f=\delta  \left(c_2^2 \Delta _H^2+2 c_2 \bar{q} \Delta _H-q_n^2+\bar{q}^2\right)^2-c_1 \Delta _H^2=0.
\]
When the system is commensurate with $q_{\mathrm{opt}}$, $f$ is negative for $\Delta_H<0$. In contrast, when $q_{\mathrm{opt}}$ is incommensurate with the system size, $f$ can be positive and the system is in the normal state. When $c_2$ is large, $q_{\mathrm{opt}}$ changes rapidly with the field, and the commensuration effect is more prominent. Meanwhile, $c_1$ determines the energy gain in the FFLO state with respect to the normal state. Therefore, for a large $c_2$ but small $c_1$, it is easier to drive the system into the normal state when $q_{\mathrm{opt}}$ is incommensurate with the system size, resulting in the re-entrant superconducting phase transition with increasing field.

\begin{figure}[t]
        \includegraphics[width=1\linewidth]{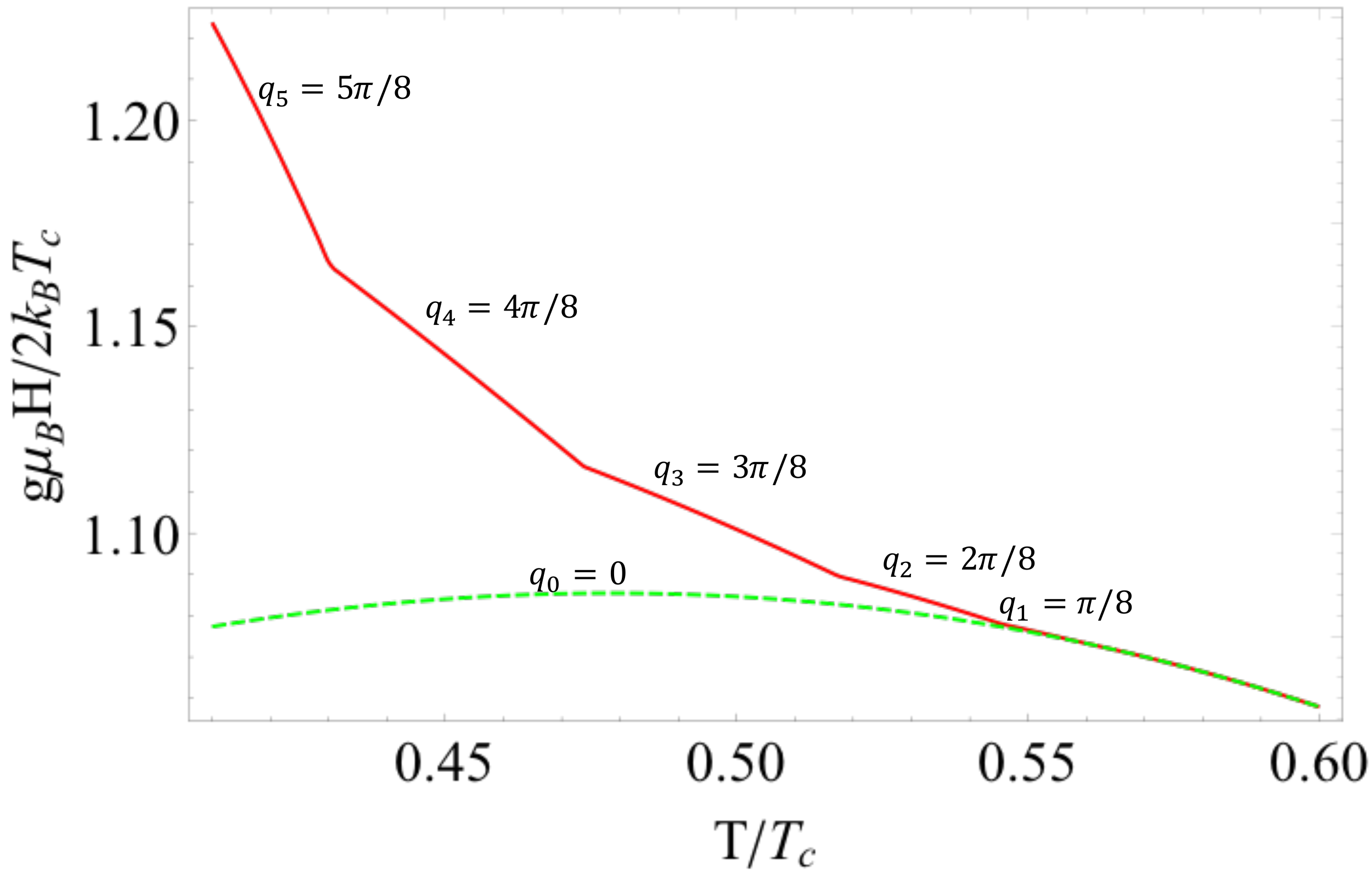}
        \caption{Superconducting phase diagram of a 1D superocnductor with $L=8\hbar v_F/k_B T_c$. The optimal $q_n$ in units of $k_B T_c/\hbar v_F$ is labeled at the phase boundary (solid line). The dashed line corresponds to the normal-superconducting phase boundary by assuming a BCS state with $q_0=0$.}
        \label{fGL2}
\end{figure}

For a parabolic band in 1D, the phase boundary is given by
\[
f=-\frac{1}{16} K_5 (q_n \hbar v_F)^4+\frac{1}{4} K_3 (q_n \hbar v_F)^2-\left(K_1-K_1^0\right)=0.
\]
The calculated phase boundary and the corresponding $q_{\mathrm{opt}}$ are depicted in Fig. \ref{fGL2}.
For this particular Fermi surface, there is no re-entrant superconducting transition as a function of the magnetic field. However, the Ginzburg-Landau approach shows that,  because of the commensuration effect, there are modulations of the phase boundary as different optimal wavevectors of the FFLO state are found.

\section{Conclusion}\label{sec7}
To summarize, we have presented a detailed study of the commensuration effect of the FFLO state in small and thin superconductors under a parallel magnetic field. We employ the Bogoliubov de-Gennes equation, Green function approach, and Giznburg-Landau theory to study the various superconducting states and phase diagrams both for the clean $s$-wave and $d$-wave superconductors. The three methods complement each other and offer a consistent picture. 

As a consequence of the strong geometry confinement, the superconducting phase diagram is strongly modulated. Moreover, there exist several FFLO pockets with different wave vectors of the order parameter separated by the normal state. Therefore, re-entrance of superconductivity by increasing the magnetic field is observable at low temperatures. Although we have restricted ourselves to a simple model, the generic features due to the commensuration effect should be valid for more realistic models relevant for experimental systems. The commensuration effect of the FFLO state can therefore offer additional evidence in experiments.

\section{Acknowledgements}
The authors thank Roman Movshovich for helpful discussions. Computer resources for numerical calculations were supported by the Institutional Computing Program at LANL. The work by SZL and TK was carried out under the auspices of the U.S. DOE Contract No. DE-AC52-06NA25396 through the LDRD program, and was supported by the Center for Nonlinear Studies at LANL.

\appendix
\section{Tridiagonal Toeplitz matrix}\label{App:A}
To obtain the Green function in the normal state, we need to use some properties of the tridiagonal Toeplitz matrices. 
For a one-dimensional lattice with open boundary condition, the hopping term of the Hamiltonian can be written as a tridiagonal Toeplitz matrix, whose eigenvectors are
\begin{eqnarray}
v^k=\sqrt{\frac{2}{N+1}}\left(\sin \left(\frac{1 \pi  k}{N+1}\right),\ldots ,\sin \left(\frac{N \pi  k}{N+1}\right)\right).
\end{eqnarray}
The corresponding eigenvalues are
\begin{eqnarray}
\lambda _{k}=2 t \cos \left({\frac {\pi k}{N+1}}\right)
\end{eqnarray}
where $k=1, \ldots , N$, and $N$ is the system size.

For a two dimensional lattice, the hopping term can be written as a Kronecker product of a tridiagonal Toeplitz matrix and the identity matrix of the same dimension. The eigenvectors are
\begin{eqnarray}
&&\sqrt{\frac{2}{N_x+1}}\left(\sin \left(\frac{1 \pi  k_x}{N_x+1}\right),\ldots ,\sin \left(\frac{N_x \pi  k_x}{N+1}\right)\right)
\nonumber\\
\otimes
&&\sqrt{\frac{2}{N_y+1}}\left(\sin \left(\frac{1 \pi  k_y}{N_y+1}\right),\ldots ,\sin \left(\frac{N_y \pi  k_y}{N_y+1}\right)\right),
\end{eqnarray}
and the eigenvalues are
\begin{eqnarray}
\lambda _{k}=2 t \cos \left({\frac {\pi k}{N+1}}\right).
\end{eqnarray}
Here $k_x=1, \ldots , N_x$ and $k_y=1, \ldots , N_y$ with $N_x\times N_y$ being the system size. Using these eigenvectors, the normal Green function with open boundary condition can be evaluated.

\section{Evaluation of the Green functions in Eqs.~\eqref{eq:self_consistent_equation_swave} and ~\eqref{eq:self_consistent_equation_dwave}}\label{App:B}
In Eqs.~\eqref{eq:self_consistent_equation_swave} and ~\eqref{eq:self_consistent_equation_dwave}, we need to evaluate the function of the form 
\begin{eqnarray}
\frac{1}{\hbar}\int d\tau  G^*_{0-- }\left(\tau  x_1,\tau _1 x_2\right) G_{0++}\left(\tau  x_3,\tau _1 x_4\right).
\end{eqnarray}
Using Eq.~\eqref{eq:normal_greens_function}, the function becomes
\begin{eqnarray}
\sum_{k_1,k_2,\omega_n} 
\frac{u_{k_1}(x_1) u_{k_1}(x_2) 
u_{k_2}(x_3) u_{k_2}(x_4)}
{\beta \left(-i \hbar\omega_n-\xi _{-}\right)\left(i \hbar\omega_n-\xi _{+} \right) },
\end{eqnarray}
where $u_k(x)=\frac{2}{\sqrt{(1+N_x)(1+N_y)}}\sin\left(\frac{\pi x^{(1)} k_x}{N_x+1}\right)\sin\left(\frac{\pi x^{(2)} k_y}{N_y+1}\right)$. We carry out the summation over $\omega_n$:

	\begin{eqnarray}
		&&\frac{1}{\beta   }\sum _{\omega_n }
		\frac{1}{\left(-i \hbar\omega_n-\xi _{-}\right)\left(i \hbar\omega_n-\xi _{+} \right) },
	\end{eqnarray}
	where $\omega_n=\left(2n+1\right)\pi/\beta\hbar$ for fermions. Using the contour integral method, the above expression is equivalent to
	\begin{eqnarray}
		&&-\frac{\hbar}{2\pi   }\oint
		\frac{1}{\left(-i \hbar z-\xi _{-}\right)\left( i \hbar z-\xi _{+} \right)}\frac{1}{e^{i \beta \hbar z}+1},
	\end{eqnarray}
	where the contour encircles the real axis. Deforming the contour to encircle the poles on the imaginary axis, the integral becomes
	\begin{eqnarray}\label{eq:twof}
		&&\frac{f\left(\frac{1}{2}\left(\xi _{+}-\xi _{-}-  |\xi _{+}+\xi _{-}|\right)\right)}{|\xi _{+}+\xi _{-}|}
        -\frac{f\left(\frac{1}{2}\left(\xi _{+}-\xi _{-}+  |\xi _{+}+\xi _{-}|\right)\right)}{|\xi _{+}+\xi _{-}|}, \nonumber \\
	\end{eqnarray}
	where $f(x)$ is the Fermi function. Using the following identity
	\begin{eqnarray}
		&&\frac{1}{e^a+1}-\frac{1}{e^b+1}
		=\frac{\sinh \left(\frac{b-a}{2}\right)}{\cosh \left(\frac{a-b}{2}\right)+\cosh \left(\frac{a+b}{2}\right)},
	\end{eqnarray}
	Eq.~\eqref{eq:twof} becomes
	\begin{eqnarray}
		\frac{\sinh (\beta \xi )}{2 \xi\left(\cosh (\beta  \zeta )+\cosh (\beta  \xi)\right)},
	\end{eqnarray}
	where  $\xi =\left(\xi _{+}+\xi _{-}\right)/2$ and $\zeta =\left(\xi _{+}-\xi _{-})\right)/2$. After applying a similar calculation to the other term with $(\sigma\leftrightarrow -\sigma)$ and collecting the terms, Eq.~\eqref{eq:self_consistent_equation_swave} becomes
    	\begin{eqnarray}
	        \Delta ^*_{x_1}
		&&	=
		V\sum_{k_1,k_2,x_2} \Delta ^*_{x_2}
\frac{\sinh (\beta \xi )}{2 \xi\left(\cosh (\beta  \zeta )+\cosh (\beta  \xi)\right)}\nonumber\\
&&
\times
u_{k_1}(x_1) u_{k_1}(x_2) 
u_{k_2}(x_1) u_{k_1}(x_2),
	\end{eqnarray}
and Eq.~\eqref{eq:self_consistent_equation_dwave} becomes
\begin{eqnarray}
	        \Delta ^{r_1*}_{x_1}
		&&	=
		V\sum_{k_1,k_2,x_2,r_2} \Delta ^{r_2*}_{x_2}
\frac{\sinh (\beta \xi )}{2 \xi\left(\cosh (\beta  \zeta )+\cosh (\beta  \xi)\right)}\nonumber\\
&&
\times
u_{k_1}(x_1) u_{k_1}(x_2) 
u_{k_2}(x_1+r_1) u_{k_1}(x_2+r_2).
	\end{eqnarray}

\bibliographystyle{apsrev}    

\begin{thebibliography}{48}
\bibitem[{Dir()}]{DirtySCNote}
	\bibinfo{note}{The superconductors in those experiments are in the dirty limit.
		One challenge would be to fabricate clean superconductors to favor the FFLO
		state.}
\end{thebibliography}

\end{document}